\documentclass[11pt,twocolumns]{IEEEtran}
\ifCLASSOPTIONcompsoc
\else
\fi
\usepackage[bibstyle=numeric-comp,citestyle=numeric-comp, uniquename=false,sortcites=true,maxbibnames=10, maxcitenames=2,doi=false, url=false,  backend=bibtex]{biblatex}
\bibliography{literature_latin1}

\usepackage{graphics, graphicx}
\graphicspath{{./figures/}}

\ifCLASSINFOpdf
   \usepackage{graphics, graphicx}
\else
\fi
\usepackage[cmex10]{amsmath}
\usepackage{amssymb}
\usepackage{mdwmath}
\usepackage{mdwtab}

\ifCLASSOPTIONcompsoc
  \usepackage[font=normalsize,labelfont=sf,textfont=sf,caption=false]{subfig}
\else
  \usepackage[font=footnotesize,caption=false]{subfig}
\fi
\ifCLASSOPTIONcompsoc
  \usepackage[caption=false,font=normalsize,labelfont=sf,textfont=sf]{subfig}
\else
  \usepackage[caption=false,font=footnotesize]{subfig}
\fi
\usepackage{url}

\usepackage{color}

\usepackage{units}
\usepackage{multirow}

\newtheorem{rema}{Remark}
        
\newtheorem{prob}{Problem}   	
\newcommand{\sign}{\operatorname{sign}}
\newcommand{\minimize}{\operatorname{minimize}}
\newcommand{\diag}{\operatorname{diag}}
\newcommand{\LWR}{KUKA LWR IV \,}

\newcommand{\mbb}[1]{\mathbb #1}
 
\newcommand{\mcl}[1]{\mathcal #1}

\newcommand{\Rnx}{\mbb{R}^{n_x}}

\newcommand{\Rnu}{\mbb{R}^{n_u}}

\begin{document}
%
\title{Implementation of Nonlinear Model Predictive Path-Following Control for an Industrial Robot}
%
%
%
%

\author{\normalsize Timm~Faulwasser~\IEEEmembership{Member,~IEEE},  Tobias Weber, Pablo Zometa, Rolf Findeisen~\IEEEmembership{Member,~IEEE} 

\IEEEcompsocitemizethanks{\IEEEcompsocthanksitem TF is with the Institute for Applied Computer Science, Karlsruhe Institute of Technology, Germany.
He is also with  the Laboratoire d'Automatique, \'Ecole Polytechnique F\'ed\'erale de Lausanne, Switzerland. 
TW is with the Institute for Mathematical Optimization, Otto-von-Guericke University  Magdeburg, Germany.  PZ and RF are with the Institute for Automation Engineering, Otto-von-Guericke University  Magdeburg, Germany. 
E-mails: timm.faulwasser@\{epfl.ch,\,kit.edu\}; \{tobias.weber, pablo.zometa, rolf.findeisen\}@ovgu.de.}
%
}

%
%

\markboth{~}%
{Shell \MakeLowercase{\textit{et al.}}: Bare Demo of IEEEtran.cls for Computer Society Journals}
%



\IEEEcompsoctitleabstractindextext{%
\begin{abstract}
Many robotic applications, such as milling, gluing,
or high precision measurements, require the precise following
of a pre-defined geometric path.
We investigate the real-time feasible implementation of model predictive path-following control for an industrial robot. 
We consider constrained output path following with and without reference speed assignment. Finally, we present results of an implementation of the  proposed model predictive path-following controller on a \LWR robot. 

%

\end{abstract}

\begin{IEEEkeywords}
path following, nonlinear model predictive control, constraints, optimal control, \LWR
\end{IEEEkeywords}}

\maketitle

\IEEEdisplaynotcompsoctitleabstractindextext

%
\IEEEpeerreviewmaketitle

\section{Introduction}

Not all control tasks arising in applications fit well into the classical framework of set-point stabilization and trajectory tracking. 
For instance, consider tasks such as steering an autonomous vehicle along a given reference track, precise machine tooling, or control of autonomous underwater vehicles. All these applications have in common that a system should be steered along a pre-specified geometric curve in a (position) output space, whereby the speed to move along the curve is not fixed a priori. Such control tasks are typically denoted as \textit{path-following problems}.

These problems arise frequently in the context of dynamic motion planning and trajectory generation for mechatronic systems including bi-pedal walking and standard robots  \cite{Wieber06,Shin85a, Westervelt07, Kumar99}. Typically, a dynamic motion is assigned to a geometric reference path by solving an optimal control problem, or via some heuristics,  leading to the desired reference motion along the path. The reference motion itself is then tracked by means of some feedback controller. In other words, path-following problems are frequently decomposed into trajectory generation and trajectory tracking, i.e., path following is reformulated as trajectory tracking.

To avoid the reformulation as a tracking problem, different closed-loop path-following control schemes have been proposed, see e.g. \cite{epfl:faulwasser15b,ifat:faulwasser13a_short, Lam10, Banaszuk95a,Nielsen10a, Aguiar05a}.
The common underlying idea of these schemes is that the generation of the reference motion along the path, and the computation of inputs to track this motion, are both done in an integrated fashion at the run-time of the controller. In other words, a closed-loop path-following controller directly modulates the reference speed along the geometric path to reduce the path-following error and, at the same time, computes inputs to track this reference motion. 
Besides geometric feedback designs for path following---e.g. \cite{Banaszuk95a,Nielsen10a, Aguiar05a}---different model predictive control approaches tailored to path following have been investigated, see \cite{ifat:faulwasser13a_short,Lam10,epfl:faulwasser15b,boeck13a}. 

 To this date, only a few laboratory implementations of model predictive control tailored to path-following problems of mechatronic and robotic systems have been reported: Discrete time predictive path-following control of an x-y table is presented in \cite{Lam11a}. Real-time implementations of sampled-data predictive path-following controllers have been presented in 
 \cite{boeck13a}, which is based on differential flatness, and \cite{epfl:faulwasser13b}, wherein predictive path following of an industrial robot with paths defined in the joint space is discussed.

\begin{figure}[t]
\begin{center} 
 \includegraphics[width=0.45\textwidth]{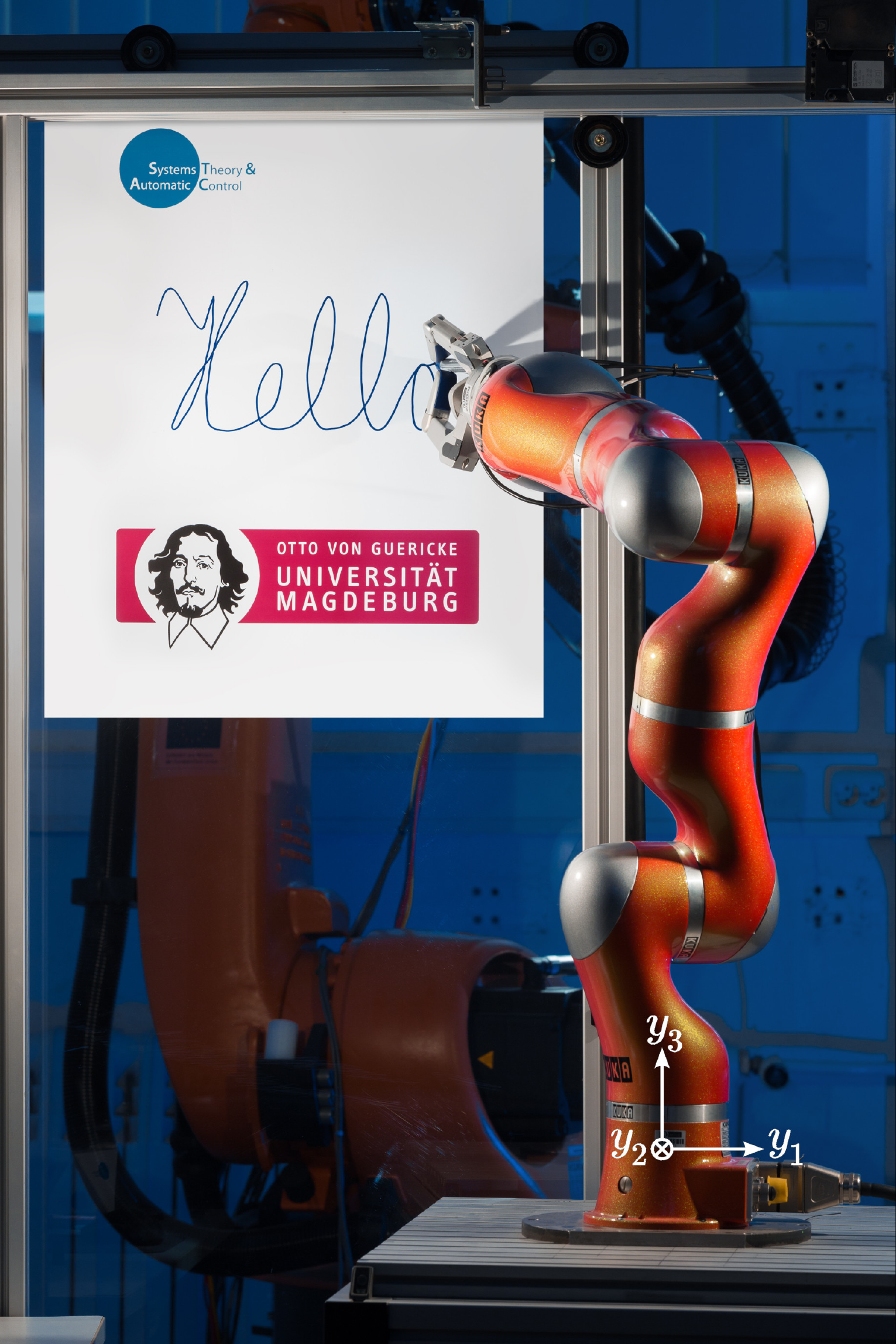} 
  \caption{Considered path-following problem: \LWR writing on a white board.\label{fig:robot_setup}}
\end{center}
\end{figure}

In the present paper, we discuss the design and implementation of a  sampled-data nonlinear model predictive path-following control scheme in the presence of input and state constraints. Our main contribution is a \textit{proof-of-concept} demonstration of predictive path following. The results are obtained from a real-time feasible implementation on a \LWR robot in a configuration with three actuated joints, see Figure \ref{fig:robot_setup}. 
The implementation relies on a predictive control approach presented in \cite{epfl:faulwasser15b}, wherein system-theoretic properties, such as stability and path convergence are analyzed, while implementation aspects and experimental results are not discussed. 
In contrast to the result presented in \cite{epfl:faulwasser13b}, we use a robot configuration with three joints instead of two. Furthermore, we consider reference paths defined directly in the operational space of the robot, i.e., the Cartesian space. 
We demonstrate that a suitable numerical implementation allows solving path-following problems achieving a sampling period in the order of \unit[1]{ms}. 
A detailed comparison of predictive path-following with predictive trajectory-tracking formulations is beyond the scope of this paper, we refer to \cite{boeck13a} for such a comparison.

The remainder of this paper is structured as follows: Section \ref{sec:MPFC} recalls the formal problems of  constrained output path following and speed-assigned path following for general nonlinear systems; additionally we sketch the conceptual ideas of model predictive path-following control used here. Details of the implementation on a \LWR are discussed in Section \ref{sec:implement}; experimental results are presented in Section \ref{sec:results}.

\  \section{Predictive Path Following}\label{sec:MPFC}
  We consider nonlinear systems of the form
  \begin{subequations} \label{eq:sys}
    \begin{align}
      \dot{x} & = f(x,u), \quad x(t_0) = x_0, \label{eq:sys_dyn}\\
      y & = h(x), \label{eq:sys_out}
    \end{align}
  \end{subequations}
  where $x \in \Rnx$, $u \in \Rnu$, and $y \in \Rnu$ represent the state,
  the input and the output.
  The states are constrained to a closed set, i.e., for all $t: x(t) \in \mcl{X} \subseteq \Rnx$.
  The inputs $u: [t_0, \infty) \to \mcl{U}$
    are piece-wise continuous and take values in a compact set $\mcl{U} \subset \Rnu$, which is briefly denoted by
    $u(\cdot) \in \mcl{P}\mcl{C}(\mcl{U})$. 
    The maps $f: \Rnx \times\Rnu\to \Rnx$ and $h: \Rnx \to \Rnu$ 
    are assumed to be sufficiently often continuously differentiable and 
    \eqref{eq:sys_dyn} is considered to be locally Lipschitz. Also note that the control system is
    assumed to have a square input-output structure, i.e., $\dim u = \dim y := n_u$. The solution of \eqref{eq:sys_dyn} at time $t$, originating at time $t_0$ from $x_0$, driven by an input $u(\cdot)$, is denoted as $x(t, t_0, x_0 | u(\cdot))$.

\subsection{Path-following Problems}
Output path-following refers to the task of tracking/following a geometric reference in the output space \eqref{eq:sys_out} \cite{Skjetne04,Nielsen10a}. Here, we assume that this reference is given by
\begin{equation} \label{eq:path}
 \mcl{P} = \left\{ y \in \Rnu ~|~ \theta \in \mbb{R} \mapsto y = p(\theta)\right\}.
\end{equation}
The scalar variable $\theta \in \mbb{R}$ is called path parameter, and $p(\theta)$ is a parametrization of $\mcl{P}$. Note that in path-following problems there is typically no strict requirement \textit{when to be where} on $\mcl{P}$. 
In other words, the path parameter $\theta$ is time dependent but its time evolution $t \mapsto \theta(t)$ is not specified a priori. Rather the system input 
$u(\cdot)$  and the timing $\theta(\cdot)$
are to be chosen such that the path is followed as exactly as possible. Furthermore, note that in some cases, such as finitely long paths, it can be helpful to restrict the path parameter to $ [\theta_0, \theta_1]$, where $\theta_0$ is the start point of the path and $\theta_1\in [\theta_0, \infty)$ denotes the end point of the path. 

Subsequently, we  consider the problem of steering the output \eqref{eq:sys_out} to the path $\mcl{P}$ and following it along in the direction
of increasing  values of $\theta$.

\begin{prob}[{Constrained output path following \cite{epfl:faulwasser15b}}] \label{prob:path} 
Given the system \eqref{eq:sys} and the reference path $\mcl{P}$ \eqref{eq:path}, design a controller  that computes $u(\cdot)$ and $\theta(\cdot)$  and achieves:
\begin{itemize}
	\item[i)] {Path convergence}: The system output $y =h(x)$ converges to the set $\mcl{P}$ in the sense that
	\[ \underset{t \to \infty}{\lim} \| h(x(t)) - p(\theta(t)) \| = 0.\]  		
	\item[ii)] {Convergence on path}: The system moves along $\mcl{P}$
	 in forward direction, i.e.
	\[ \dot\theta(t) \geq 0 \quad \textrm{and}\quad \underset{t \to \infty}{\lim} \| \theta(t) -\theta_1 \| = 0.\]  	
	\item[iii)] {Constraint satisfaction}:  The constraints on the states $x(t) \in \mcl{X}$ and the inputs
	 $u(t) \in \mcl{U}$ are satisfied for all times. \hfill $\square$
\end{itemize}
\end{prob}

Instead of this formulation, one might demand that the path parameter velocity $\dot \theta(t)$ converges to a pre-specified evolution $\dot\theta_{ref}(t)$, cf. \cite{Skjetne04}. This leads to the following problem.

\begin{prob}[{Speed-assigned path following \cite{epfl:faulwasser15b}}] \label{prob:path_velo} 
Given the system \eqref{eq:sys} and the reference path $\mcl{P}$ \eqref{eq:path}, design a controller that computes $u(\cdot)$ and $\theta(\cdot)$, achieves part i) \& iii) of Problem \ref{prob:path} and guarantees:
\begin{itemize}
	\item[ii)] {Velocity convergence}: The path velocity $\dot\theta(t)$ converges to a predefined
	profile such that
	\[ \underset{t \to \infty}{\lim} \| \dot\theta(t) -\dot\theta_{ref}(t) \| = 0. \vspace*{-.7cm}\]  	
	\hfill $\square$	
\end{itemize}
\end{prob}
\vspace*{2mm}
\begin{rema}[Speed-assigned paths and tracking]
Note that path following with velocity assignment is not equivalent to trajectory tracking, as speed assignment does not specify a unique output reference $p(\theta(t))$. Rather, the problem with speed assignment admits several reference trajectories  $p(\theta_i(t)),  i   \in\{1,2,\dots\}$, with $\dot\theta_i(t) =\dot\theta_{ref}(t)$ differing with respect to $\theta$, i.e., $\theta_i(t) \neq \theta_j(t),\, i\neq j$. Furthermore,  in case disturbances lead to large path deviations, path following with speed assignment allows adjusting the timing $\theta(t)$, while this can be challenging in standard trajectory-tracking formulations. \hfill $\square$
\end{rema}

\vspace*{2mm}
The conceptual idea of  path following is to treat the path parameter $\theta$ as a virtual state whereby the time evolution $t \mapsto \theta(t)$
is  influenced by an extra input, cf. \cite{Skjetne04, Aguiar05a, ifat:faulwasser13a_short,epfl:faulwasser15b}. Usually, the time evolution $t \mapsto \theta(t)$ is described by an additional differential equation termed \textit{timing law}. Basically, the timing law is an extra degree of freedom in the controller design.  Subsequently, we rely on a simple integrator chain as timing law 
\begin{equation} \label{eq:timing}
\theta^{(\hat r +1)} = v,
\end{equation}
 where $\hat r \in \mbb{N}$ is sufficiently large as outlined in Remark \ref{rem:TNF}. We note that more complex timing laws can be considered.
The \textit{virtual} input of the timing law is assumed to be piece-wise continuous and bounded, i.e.,  $v(\cdot) \in \mcl{P}\mcl{C}(\mcl{V}), 
\mcl{V} \subset \mbb{R}$.
Similar to \cite{ifat:faulwasser13a_short, epfl:faulwasser15b}, we use the compact notation $z := (\theta, \dot \theta, \dots, \theta^{(\hat r)})^T$ of \eqref{eq:timing} to formulate path-following problems via the augmented system 
\begin{subequations} \label{eq:sys_aug}
\begin{align}
 \begin{pmatrix}
    \dot x \\ \dot z 
  \end{pmatrix} &=  \begin{pmatrix}
				f(x,u) \\
				l(z,v)
		     \end{pmatrix},  \quad  \begin{pmatrix}x(t_0) \\ z(t_0) \end{pmatrix} =  \begin{pmatrix}x_0 \\ z_0 \end{pmatrix}, \label{eq:aug_state}\\
  \begin{pmatrix}
    e \\ z
  \end{pmatrix} &= \begin{pmatrix}
			h(x) - p(z_1) \\
			z
  \end{pmatrix}. \label{eq:aug_output}
\end{align}
\end{subequations}
The output \eqref{eq:aug_output} consists of two elements, the path following error $e = h(x) - p(\theta)$, and the full virtual state  $z$.
With respect to the augmented system \eqref{eq:sys_aug} output path-following (Problem \ref{prob:path}) requires that the error $e$ converges to zero while the path parameter $\theta$ converges to $\theta_1$---the final path point. 

\begin{rema}[{Choice of suitable timing laws}]  \label{rem:TNF} ~\\
How should one choose the parameter $\hat r$ in the timing law \eqref{eq:timing}?
This question can be answered using tools from geometric nonlinear control and the concept of \textit{transversal normal forms}, see \cite{Banaszuk95a, Nielsen08a, ifat:faulwasser13a_short, epfl:faulwasser15b}. 
The main idea is to choose $\hat r$ sufficiently large such that one can map the augmented system \eqref{eq:sys_aug} at least locally into suitable coordinates allowing for characterization of the manifold on which any state trajectory corresponds to an output trajectory traveling along $\mcl{P}$. \hfill $\square$
\end{rema}

\subsection{Model Predictive Path-following Control}
We tackle output path-following problems with and without speed assignment in presence of input and state constraints (Problems \ref{prob:path} \& \ref{prob:path_velo}) via a continuous time sampled-data nonlinear model predictive control (NMPC) scheme, denoted as model predictive path-following control (MPFC), \cite{ifat:faulwasser13a_short,epfl:faulwasser15b}.

MPFC is based on the augmented system description \eqref{eq:sys_aug}.  As common in model predictive control, the system input is obtained via the repetitive solution of an optimal control problem (OCP). At each sampling instance $t_k = t_0 +
k\delta$, with $k \in \mbb{N}_0$ and sampling period $\delta > 0$, the cost functional to be minimized is
\begin{multline} \label{eq:J}
  J\left(x(t_k), z(t_k), \bar u(\cdot), \bar v(\cdot)\right)   \\ = \int_{t_k}^{t_k +T}
  F\left(\bar e(\tau), \bar z(\tau),  \bar u(\tau),
  \bar v(\tau) \right)d\tau. 
\end{multline}
As usual in NMPC, $F:  \Rnu\times\mcl{Z}\times\mcl{U}\times\mcl{V}\to \mbb{R}_0^+$ is
called cost function and $T \in (\delta, \infty)$ denotes the prediction horizon. 
The OCP solved repetitively is:
\begin{subequations} \label{eq:OCP}
  \begin{equation}
    \underset{(\bar u(\cdot), \bar v(\cdot))\in \mcl{P}\mcl{C}(\mcl{U}\times\mcl{V})}{\minimize} ~   J\left(x(t_k), z(t_k), \bar u(\cdot), \bar v(\cdot)\right)  
  \end{equation}
  subject to the constraints
  \begin{align}
 \dot{\bar{x}}(\tau)& = f(\bar x(\tau),\bar u(\tau)), \quad \bar x(t_k) =
    x(t_k)  \label{eq:con_x}\\
    \quad \dot{\bar{z}}(\tau)& = l(\bar z(\tau),\bar v(\tau)), \quad~ \bar{z}(t_k) = z(t_k) \label{eq:con_z}\\
    \quad \bar e(\tau) &= h(\bar x(\tau)) - p(\bar z_1(\tau)) \label{eq:con_e} \\
    \quad \bar x(\tau) &\in \mcl{X}, ~\bar 	u(\tau)  \in \mcl{U}  \label{eq:con_X}\\
    \quad \bar z(\tau) &\in \mcl{Z}, ~\bar 	v(\tau)  \in \mcl{V} \label{eq:con_Z} 
  \end{align}
\end{subequations}
which have to hold for all  $\tau \in [t_k, t_k + T]$. As common in NMPC we denote predicted variables, i.e., internal variables of the controller, by superscript $\bar\cdot$.
The MPFC scheme \eqref{eq:OCP} is built upon the augmented dynamics \eqref{eq:sys_aug}, and thus they are considered as dynamic constraints in \eqref{eq:con_x}--\eqref{eq:con_z}.

At the core of the MPFC scheme is the repeated solution of \eqref{eq:OCP} in a receding horizon fashion at all sampling instants $t_k$.
The solution to \eqref{eq:OCP} are optimal input trajectories,
denoted as $\bar u^{\star}(\cdot, x(t_k))$ and $\bar v^{\star}(\cdot, z(t_k))$.   
Solving \eqref{eq:OCP} at time $t_k$ with finite horizon $[t_k, t_k+T]$, one plans a reference motion $t \mapsto p( \theta(t)) \in \mcl{P}$ and, at the same time,  computes the system inputs to track these trajectories. 
Finally, $\bar u^{\star}(\cdot, x(t_k))$ is applied to system \eqref{eq:sys}
\[ \forall t \in [t_k, t_k + \delta): \quad u(t) = \bar
u^{\star}(t, x(t_k)).
\]
At the next sampling instant $t_{k+1} = t_k + \delta$,  the OCP \eqref{eq:OCP} is solved again for new initial conditions.
While at each sampling instance the measured or observed state 
$x(t_k)$ serves as initial condition for \eqref{eq:con_x}
the initial conditions of the  timing law \eqref{eq:con_z} is the last predicted trajectory
evaluated at time $t_k$, i.e. $z(t_k) = \bar{z}(t_k, t_{k-1},\bar{z}(t_{k-1})|\bar v_{k-1}(\cdot)).
$\footnote{
If no initial condition for the first sampling instance $k=0$ is given, 
one can use $z(t_0)=  (\theta(t_0), 0, \dots, 0)^T$ whereby $\theta(t_0)$ locally minimizes the distance 
$\| h(x_0) -p(\theta)\|$.}

Note that in real-time feasible implementations of predictive control schemes one will typically compute an approximation of the optimal solution $\bar u^{\star}(\cdot, x(t_k))$, i.e., one will apply a feasible but sub-optimal iterate of a numerical solution scheme.
It is worth mentioning that the MPFC scheme \eqref{eq:OCP} does not aim at a time-optimal 
motion along $\mcl{P}$ as it is often considered in robotics \cite{Shin85a,Slotine89a}. 
Rather, we aim at a feedback strategy ensuring that the system output converges to the path and moves along the path in forward direction.

Summarizing, the MPFC scheme is built upon the receding horizon solution of \eqref{eq:OCP}, whereby the system model \eqref{eq:con_x} with the real system input $u$ \textit{and} the virtual path parameter dynamics \eqref{eq:con_z} with the virtual input $v$ are part of the optimization problem. 
Since, the initial condition of \eqref{eq:con_z} at time $t_k$ is based on the previous solution at time $t_{k-1}$, the variable $z$ can be understood as an internal state of the controller. In other words, the MPFC scheme based on \eqref{eq:OCP} is a dynamic feedback strategy, cf. \cite[Remark 1]{epfl:faulwasser15b}.

\begin{rema}[{Convergence conditions for MPFC}] 

It is fair to ask for conditions ensuring that the proposed MPFC scheme solves Problems \ref{prob:path} \& \ref{prob:path_velo} or guarantees path convergence. The present paper focuses on the application and implementation of the MPFC scheme. Thus a detailed investigation is beyond its scope.
As discussed in \cite{ifat:faulwasser13a_short, epfl:faulwasser15b}, one can  establish sufficient conditions by adding an end penalty and a terminal constraint to the OCP \eqref{eq:OCP}. This way one can ensure path convergence and recursive feasibility in the presence of state constraints. 
 \hfill $\square$
\end{rema}

\subsection{Problems with and without Speed Assignment}\label{ssec:F}
So far we have put the focus on the general MPFC formulation. Next, we show how to account for problems with and without speed assignment. 

We consider a quadratic cost function $F$
\begin{multline} \label{eq:F}
F(e, z, u, v) =\\ \|(e, z_1-\theta_1, z_2-\dot{\theta}_{ref})\|^2_{Q} 
+  \left\|(u, v)\right\|^2_R,
\end{multline}
since this allows efficient computation of approximations of the Hessian of OCP \eqref{eq:OCP}. 
The weighting matrix $Q$ is positive semi-definite and $R$ is positive definite while both are diagonal, i.e., $Q = \diag(w_e, w_e, w_e, w_{\theta}, w_{\dot\theta})$ and
$R = \diag(r_u, r_u, r_u, r_v)$.  In order to converge to the path, one usually chooses $w_e\gg w_{\theta, \dot\theta}$. 
If the task at hand is a path-following problem \textit{without} speed assignment---i.e., Problem \ref{prob:path}, and one wishes to stop at $z_1 = \theta_1$---one penalizes  $z_1-\theta_1$. Hence, $w_\theta >0$ and $w_{\dot\theta}=0$ are used in this case. 
If, however, the task at hand is a path-following problem \textit{with} speed assignment---i.e., Problem \ref{prob:path_velo}, and achieving $z_2-\dot{\theta}_{ref} \approx 0$ is of interest---one uses $w_{\dot\theta}>0$ and $w_\theta =0$.
Besides the parameters of the cost function $F$ also the constraints $\mcl{Z}$ in \eqref{eq:con_Z} differ for Problems \ref{prob:path} and \ref{prob:path_velo}.
While the former problem calls for $ \mcl{Z} =  [\theta_0, \theta_1] \times [0, \infty) \times \mbb{R}^{\hat r -1}$, in the latter problem one should use 
 $\mcl{Z} =  [\theta_0, \infty) \times [0, \infty) \times \mbb{R}^{\hat r -1}$.

\section{Implementation} \label{sec:implement}

\subsection{Robot Model and MPFC Design}
For simplicity, we consider a robot configuration in which only joints 1, 2 and 4 of the \LWR are operated while the other joints are kept fixed. 
The dynamic model of the robot is given by
\begin{equation} \label{eq:robot_q}
B(q)\ddot q + C(q, \dot q)\dot q + \tau_F(\dot q) + g(q)  =  \tau.
\end{equation}
Here, $q = (q_1, q_2, q_4)^T$ is the vector of joint angles. The time derivatives $\dot q$ and $\ddot q$, respectively, refer to angular velocities and angular accelerations. The vector $\tau = (\tau_1, \tau_2, \tau_4)^T$ denotes the actuation torques applied to the coressponding joints; $B(q) =B(q)^T>0$ is the inertia matrix; $C(q, \dot q)$  represents the centrifugal and Coriolis effects. The vectors $\tau_F(\dot q)$ and $g(q)$ describe torques in the joints due to friction and gravity, respectively. Note that this model describes the robot moving freely, i.e., contact forces are not explicitly included. They are treated as disturbances. The parameters of the model can be found in \cite{ifat:bargsten13a}.

First, we rewrite the model in an implicit  state-space representation with $x_1 = q, x_2 = \dot q, u = \tau$.
This leads to 
\begin{align*}
E\begin{pmatrix} 
\dot x_1 \\  \dot x_2
\end{pmatrix}
&=
\begin{pmatrix}
x_2 \\
 u - C(x_1, x_2)x_2 - g(x_1) - \tau_F(x_2)
\end{pmatrix} \\
y &= h_{ca}(x_1), 
\end{align*}
where $E = \diag(I, B(x_1))$ is blockdiagonal.
Note that the output $y = h_{ca}(x_1)$ describes the position of the tip of the robotic arm in a Cartesian coordinate system centered at the base of the robot 
(the operational space of the robot), cf. Fig. \ref{fig:robot_setup}.  We use this output instead of the flat output $\tilde y = x_1 = q$, as path-following tasks are usually formulated in the operational space.  
Finally, to obtain the augmented system description \eqref{eq:sys_aug}, we choose  an integrator chain of length two for \eqref{eq:timing}.

We express the parametrization $p(\theta)$ of the paths as a set of polynomial splines. 
To this end, we consider an equi-distant partition of $[\theta_0, \theta_1]$ given by $\tilde\theta_i = \Delta_\theta + \tilde\theta_{i-1}, \tilde\theta_0 = \theta_0$. We describe $\mcl{P}$ by
\[
p(\theta) = \sum_{i=1}^{N_\mcl{P}} H\left(\theta- \tilde\theta_i\right)H\left(\tilde\theta_{i-1} -\theta\right)\sum_{j=0}^{N_o} a_{i,j}\theta^j.
\]
Here, $H: \mbb{R} \to \{0,1\}$ denotes the Heaviside step function and $a_{i,j}$ are polynomial coefficients. The constants $N_\mcl{P}$ and $N_o$ denote the number of path segments and, respectively, the order of the polynomial on each segment. The coefficients $a_{i,j}$ are computed such that $p(\theta)$ is continuously differentiable.

\subsection{Interfacing and Real-Time Feasible Optimization}
The considered \LWR robot arm \cite{bischoff10} is operated via the \textit{Fast Research Interface} by an external computer via an Ethernet connection \cite{Schreiber10}. This interface allows sampling rates up to \unit[1]{kHz}, which is also the sampling rate of the internal control layer \cite{bischoff10}. Furthermore, the interface allows to superpose torques on each joint when operated in  the so-called \textit{joint-specific impedance control mode}. In this mode, the torques commanded to the \LWR are composed of torques computed inside the motion kernel (i.e., gravity terms) and the torques computed by an external controller (MPFC) and transferred via Ethernet to the robot. The MPFC scheme is implemented on an external PC workstation running a Linux operating system and a Intel Xeon X5675 CPU with $3.07$ GHz clock frequency. The proposed MPFC scheme is written entirely in $C/C{++}$.

   For sake of faster computations, we simplified the model \eqref{eq:robot_q}. First, note that the friction term $\tau_F(\dot q)$ in \eqref{eq:robot_q} includes a $\sign$ function due to Coulomb friction, which is approximated by an $\arctan$ in the OCP.  Second, we rely on internal functionalities of the robot allowing for gravity compensation. Thus, the term $g(q)$ in \eqref{eq:robot_q} is neglected in the OCP. 
 The OCP \eqref{eq:OCP} is solved repeatedly at the run-time of the controller using the
automatic code generation features presented in \cite{houska2011auto}. 
Specifically, we use a direct single-shooting implementation available in version \textit{1.2.1beta} of the ACADO Toolkit. In each iteration we perform one SQP iteration, i.e., we employ a so-called real-time iteration scheme \cite{houska2011auto, Diehl01a}. We use an implicit Gauss-Legendre integrator of order 2 with 10 steps.

The prediction horizon is $T = $\unit[100]{ms} and the sampling period of the MPFC scheme is $\delta = $\unit[1]{ms}.
It is worth noting that, in contrast to discrete-time formulations of NMPC, in the sampled-data framework of  Section \ref{sec:MPFC} the choice of the input parametrization is independent of the chosen sampling period. Here, the input signals are approximated as piece-wise constant functions with 10 equi-distant intervals of $\unit[10]{ms}$.  Hence, at each sampling time $t_k$, we solve one SQP iteration with a Hessian of size $40 \times 40$.

The research interface of the robot allows to obtain the joint angles $x_1$ from magneto-resistive encoders but not the joint angular velocities $x_2$.
We employ finite differences and low pass filtering of $x_1$ to obtain angular velocities $x_2$. Note that the state $z = (\theta, \dot \theta)^T$ is merely an internal variable of the controller; thus it does not need to be estimated.

The maximum time to solve the OCP \eqref{eq:OCP} is \unit[0.48]{ms} (mean \unit[0.24]{ms}, median \unit[0.18]{ms}). The overall latency, which consists of the time to solve the OCP and the time needed for state estimation, communication, etc., is below \unit[0.92]{ms} (mean \unit[0.42]{ms}, median \unit[0.42]{ms}). Hence, in experiments a sampling rate of \unit[1]{kHz} can be achieved, which corresponds to the fastest sampling rate available via the research interface.

\section{Experimental Results} \label{sec:results}
 \vspace*{2mm}

\begin{figure}
\begin{center}
\subfloat[From top: torques, angular velocity, Cartesian path error.]{ \label{fig:clover_exp1} \includegraphics[width=0.48\textwidth]{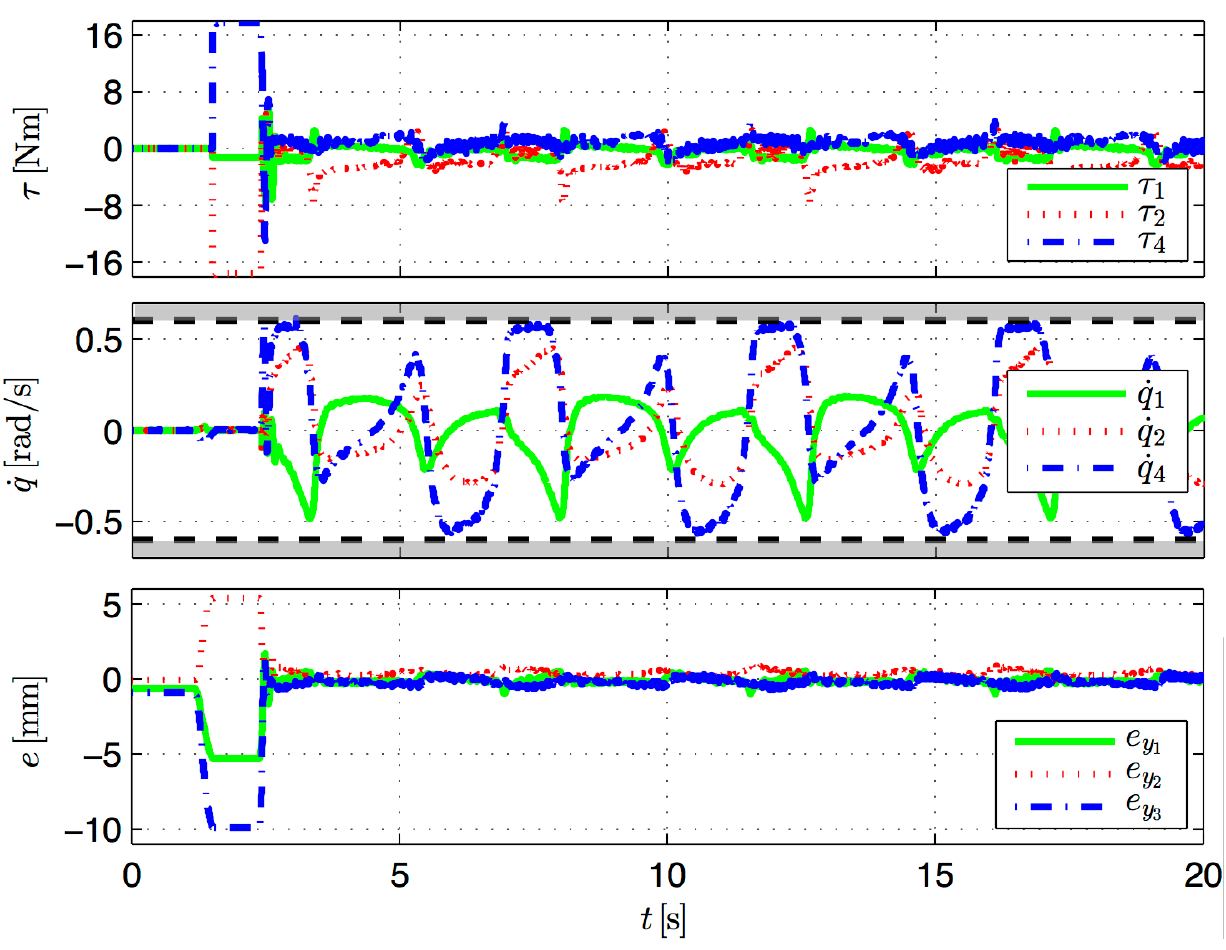}}\\
\subfloat[Virtual system.]{ \label{fig:virtual_system_tlc}  \includegraphics[width=0.48\textwidth]{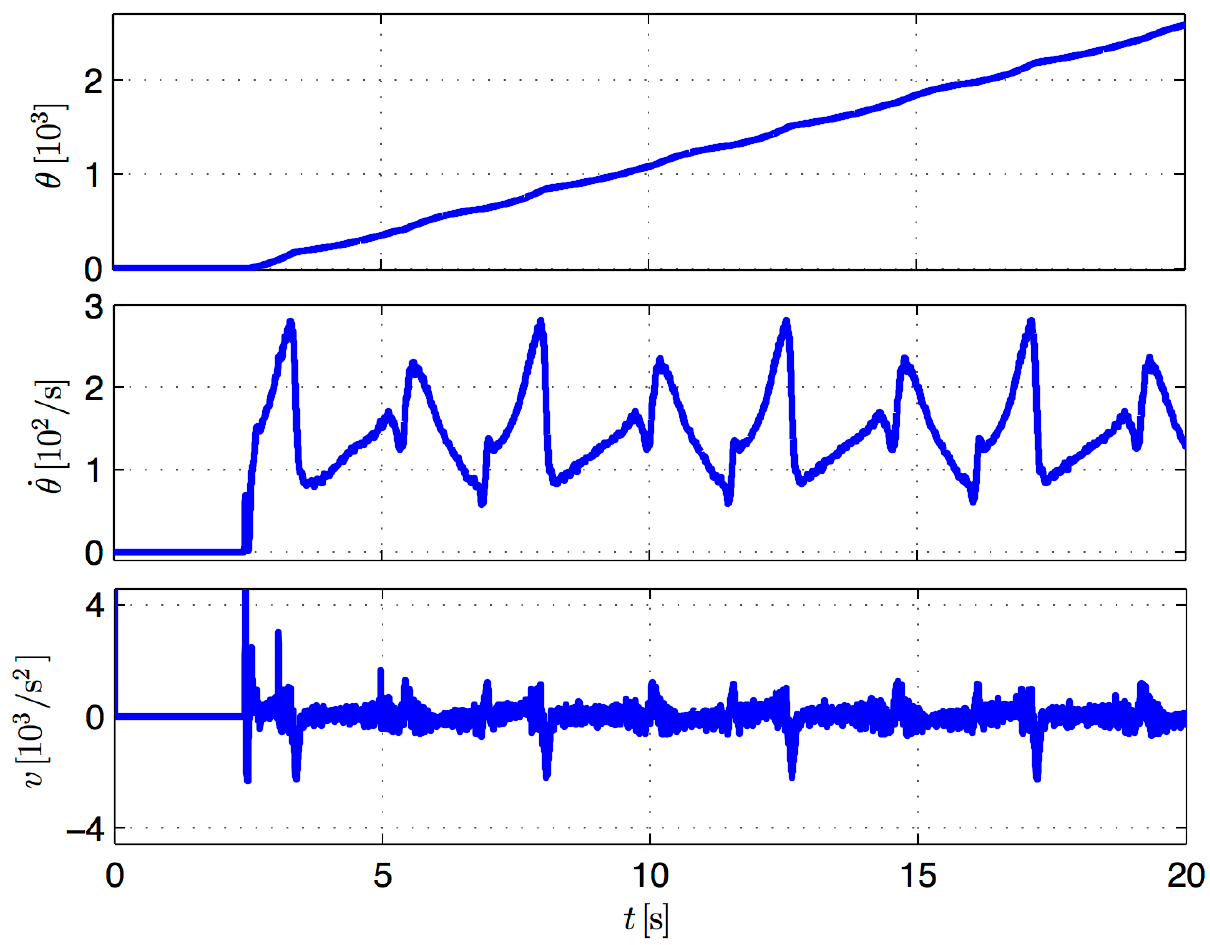}}\\
\subfloat[Cartesian position and reference path (1st turn).]{ \label{fig:cartesian_output_tlc}
 \includegraphics[width=0.52\textwidth]{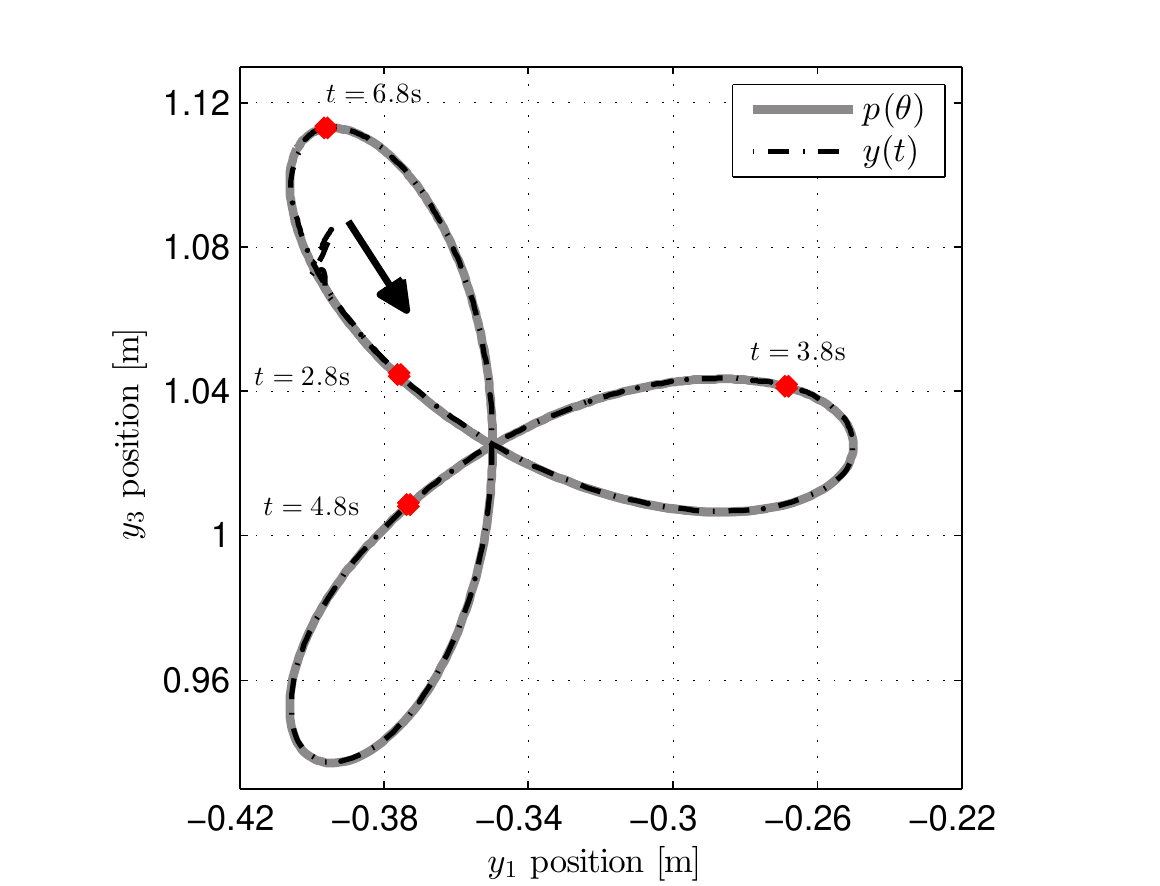}}
\caption{Experimental results for the three-leaved clover path. } \label{fig:clover}
\end{center}
\end{figure}

Two different experiments of drawing paths to a white board are presented: a three-leaved clover path and a path representing the word \textit{Hello}. The behavior of the robot during the experiments is documented in the videos available at \cite{ifat:mpfc_videos14a}. Note that the experiments shown in these videos correspond to the results depicted in Figures \ref{fig:clover}--\ref{fig:hello}. The tuning parameters of the MPFC scheme are documented in Table \ref{tab:parameters} in the Appendix.

\vspace*{2mm}
\underline{Clover Path:}\, 
The experimental results for the clover path with speed assignment are shown in Figure \ref{fig:clover}. We plot trajectories that correspond to 3 turns on the clover. 

The clover path is a closed curve, the speed assignment along the path leads to periodic behavior of the inputs and the angular velocities, see Figure \ref{fig:clover_exp1}.
This behavior can be also observed for the virtual path parameter states in Figure \ref{fig:virtual_system_tlc} as the speed along the path is  depending periodically on the curvature of the path. Furthermore, the MPFC schemes accelerates along straight parts of the path while it slows down at sharp corners of the path due to increased curvature.
Although we consider path following with speed assignment, it can be seen in Figure \ref{fig:virtual_system_tlc} that the path parameter velocity $z_2 = \dot \theta$ does not track its reference value $\dot\theta_{ref} = \unit[250]{s^{-1}}.$ The reason for this behavior is the constraint on the angular velocities of the robot, as the bound of the fourth joint is reached, see Figure \ref{fig:clover_exp1}.

 The path-following errors depicted in Figure \ref{fig:clover_exp1} are computed via the available kinematic model and the measured joint angles, i.e.
 $ e_{yi}(t) = h_{ca,i}(q(t)) - p_i(\theta(t)),\, i = 1,2,3.$
   As can be seen in Figure \ref{fig:clover_exp1}, the initial error of the experiment was rather large especially in the directions $y_2$ and $y_3$. The jump of the error at about $t =\unit[1]{s}$ is due to switching from position control to gravity compensation of the internal robot control. This gravity compensation cannot compensate exactly the contact forces between pen and board, and therefore the robot arm moves away from the starting point reached before by position control. As soon as the MPFC scheme is switched on at $t=$ \unit[2.5]{s}, the errors $e_{y_1}, e_{y_2}, e_{y_3}$ decrease rapidly and stay below \unit[1]{mm} for the rest of the experiment, see Figure \ref{fig:clover_exp1}.
The behavior of the robot in the operational space is shown in Figure \ref{fig:cartesian_output_tlc}. As one can see the MPFC scheme compensates for the initial path deviation rapidly and follows the path accurately. 

 \begin{figure}
\begin{center}
\subfloat[From top: torques, angular velocity, Cartesian path error.]{\includegraphics[width=0.48\textwidth]{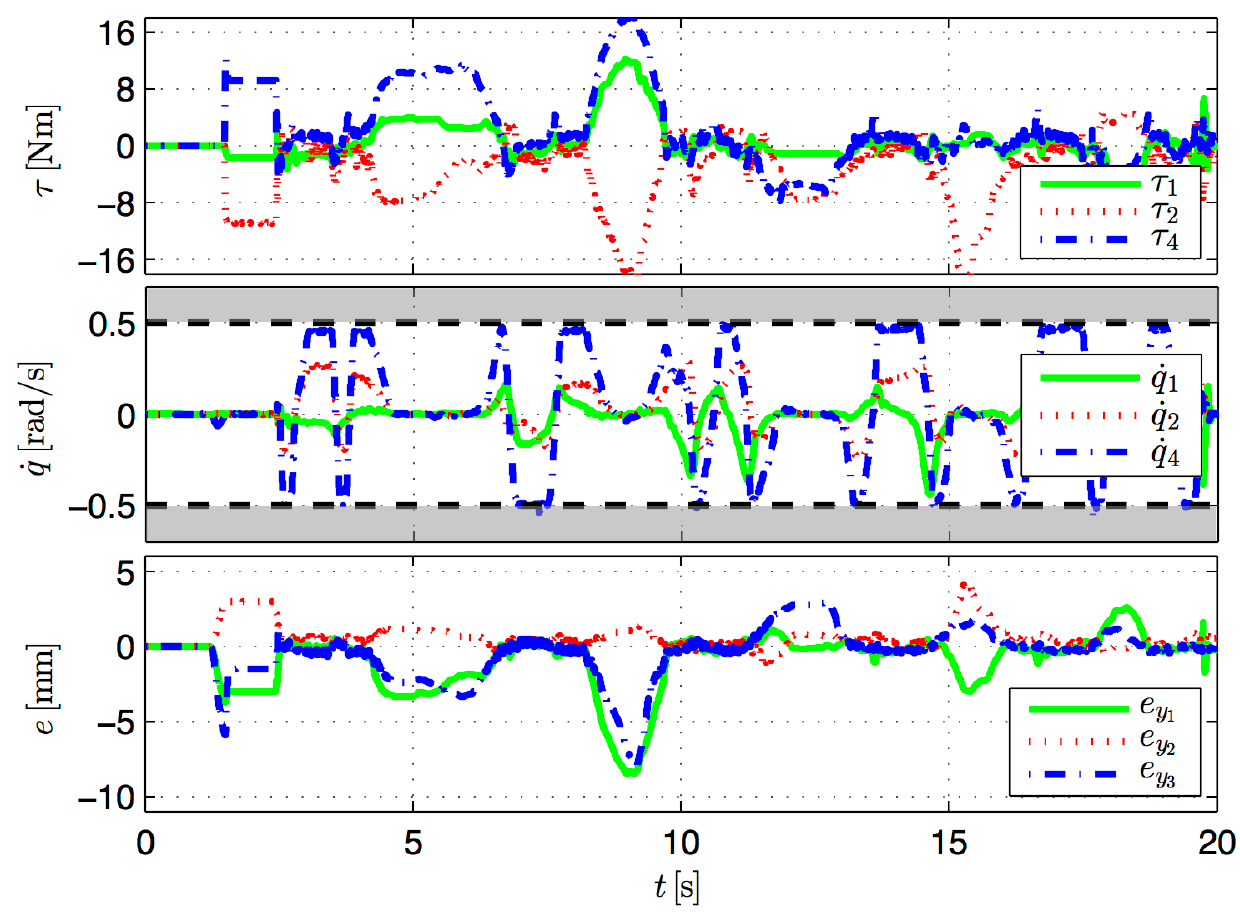} \label{fig:hello_exp_error}} \\
\subfloat[Virtual system.]{\includegraphics[width=0.48\textwidth]{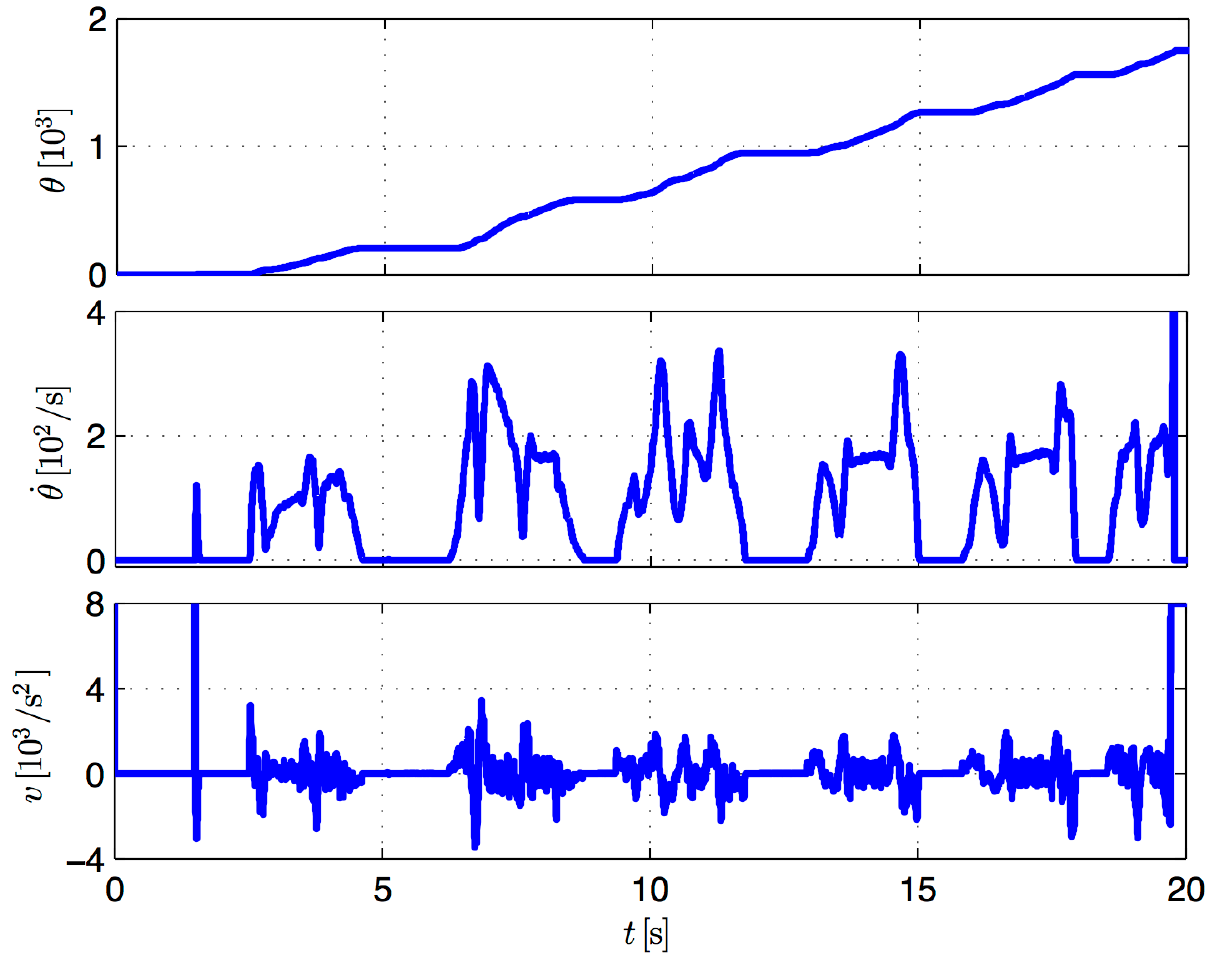} \label{fig:virtual_system_hello}} \\
\subfloat[Cartesian position and reference.]{\includegraphics[width=0.49\textwidth]{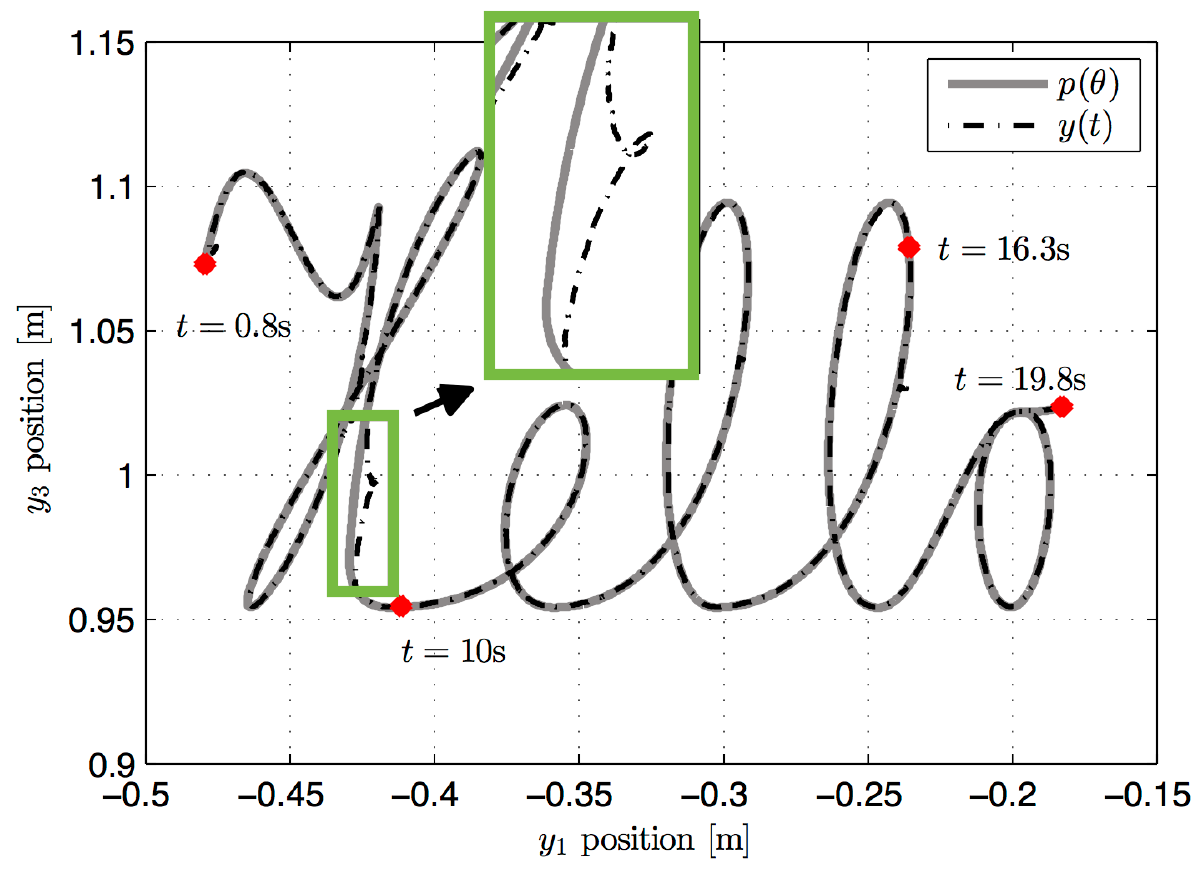}
 \label{fig:cartesian_output_hello}}
\caption{Experimental results for the \textit{Hello} path. } \label{fig:hello}
\end{center}
\vspace{10mm}
\end{figure}
\vspace*{2mm}
\underline{Hello Path:}\,
The experimental results for the \textit{Hello} path are shown in Figure \ref{fig:hello}.
During this experiment we introduced additional disturbances (external forces) by grabbing and holding the robot arm for a short time span. We applied several disturbances during five different intervals  ($[4.5,\,6.5]$, $[8.5,\,9.5]$, $[11.5,\,13]$, $[15,\,16]$, $[18,\,18.5]$).
The forces applied to the arm change the behavior of the robot completely by stopping it on the path or even moving it away from the path. This leads to large path deviations, see Figure \ref{fig:hello_exp_error}. The disturbance ends by releasing the arm and the MPFC scheme steers the robot tip back to the path. Note that during the disturbance intervals the controller is not switched off. 
The controller reacts to these disturbances (and to the increased path error) by slowing down and even stopping the reference motion along the path. This can be seen in the plots of the states and input of the virtual system in Figure \ref{fig:virtual_system_hello}. When the disturbance ends, i.e., when the external force is no longer applied, the robot tip returns and follows the path as the virtual system speeds up again.
In the zoomed-in part of Figure \ref{fig:cartesian_output_hello}, which corresponds to the disturbance during $t\in [8.5,\,9.5]$,  one can see how the robot tip is forced to move away from the reference during this interval and how it returns as soon as the disturbance ends. 
 The controller stops at the end of the path as for the \textit{Hello} path $z_1(t)-\theta_1$ is penalized. 

Note that the experiments have been performed using essentially the same parameters. The only difference is in the choice of  $w_{\theta}$ and $w_{\dot{\theta}}$ in $Q$ \eqref{eq:F} and different box constraints, cf. Table \ref{tab:parameters}. These values are passed to the optimization code during run-time, i.e. re-compilation of the auto-generated $C/C{++}$ code is not necessary.

\section{Summary and Conclusions} \label{sec:conclusion}
This paper presented results on the design and implementation of continuous time nonlinear model predictive control schemes tailored to constrained path-following problems  for robotic manipulators. We considered constrained output path following with and without speed assignment. 
We demonstrated that one can tackle both problems by changing only a few parameters of the model predictive path-following control scheme.  The real-time feasibility of the proposed scheme was illustrated via a laboratory implementation on a \LWR robot.  

The proposed model predictive path-following controller is based on an augmented system description of path-following problems allowing for direct consideration of paths defined in Cartesian space as well as input and state constraints. The presented results underpin that the proposed concept is real-time feasible and shows very promising control performance.

\section*{Appendix}
\begin{table}[h]
\caption{Implementation parameters.\label{tab:parameters}}
\vspace*{-.5cm}
\begin{center}
\begin{tabular}{| l | l | l |}
 \hline & \textit{Hello} path & clover path\\
\hline

Path segments $N_\mcl{P}$ & 1800 & 2700\\

\cline{1-3}

\multirow{ 2}{*}{$Q =\diag(q)$} & \multicolumn{2}{c|}{ 
$w_e = 10^7$ $\phantom{\Big(}$\hspace*{-2mm}}  \\
\cline{2-3}
\multirow{ 2}{*}{$q=(w_e, w_e, w_e, w_\theta, w_{\dot\theta})$} & $w_\theta=3\cdot10^{-4}$ $\phantom{\Big(}$\hspace*{-2mm} & $w_\theta = 0$\\
 & $w_{\dot\theta} = 0$ & $w_{\dot\theta} = 3\cdot10^{-4}$ $\phantom{\Big(}$\hspace*{-2mm}\\
\cline{1-3}

\cline{2-3}

$R = \diag(r_u, r_u, r_u, r_v)$ & \multicolumn{2}{c|}{ $r_u = 0.5,\, r_v = 10^{-7}$ $\phantom{\Big(}$\hspace*{-2mm} } \\

\cline{1-3}

Ref. values: $\theta_1$, $\dot{\theta}_{ref}$ $\phantom{\Big(}$\hspace*{-2mm}& $\theta_1 = 1750$ & $\dot{\theta}_{ref} = \unit[250]{s^{-1}}$ \\

\cline{1-3}

$\mcl{X} = [-\overline{x}, \overline{x}]$ & $\overline{q} = \unit[\infty]{rad}$ $\phantom{\Big(}$\hspace*{-2mm} & $\overline{q} = \unit[\infty]{rad}$  \\

$\overline{x} =  (\overline{q}, \overline{q}, \overline{q}, \overline{\dot{q}}, \overline{\dot{q}}, \overline{\dot{q}})^T$ & $\overline{\dot{q}} = \unit[0.5]{rad/s}$ & $\overline{\dot{q}} = \unit[0.6]{rad/s}$ \\

\cline{1-3}

$\mcl{U} = [-\overline{u},\overline{u}]$ $\phantom{\Big(}$\hspace*{-2mm} & \multicolumn{2}{c|}{\multirow{ 2}{*}{ $\overline{\tau}= \unit[60]{Nm}$ }} \\

$\overline{u} = (\overline{\tau}, \overline{\tau}, \overline{\tau})^T$ & \multicolumn{2}{c|}{} \\

\cline{1-3}

$\mcl{Z} = [\underline{z}, \overline{z}]$, $\underline{z} = (\theta_0, 0)^T$$\phantom{\Big(}$\hspace*{-2mm} & $\theta_0 = 0$ & $\theta_0 = 0$ \\

$\overline{z} = (\theta_1,\infty)^T$ & $\theta_1 = 1750$ & $\theta_1 =\infty$ \\

\cline{1-3}

$\mcl{V}$ & \multicolumn{2}{c|}{ $[-10^4,  8\cdot10^3]$ $\phantom{\Big(}$\hspace*{-2mm} } \\
\hline
\end{tabular}
\end{center}
\end{table}

\printbibliography

\end{document}